\documentclass[onecolumn,showpacs,preprintnumbers,amssymb]{revtex4}
\usepackage{graphicx}% Include figure files
\usepackage{dcolumn}% Align table columns on decimal point
\usepackage{bm}% bold math

\begin{document}

\title{Conductance distribution in disordered quantum wires:
crossover between the metallic and insulating regimes.}

\author{V\'{\i}ctor A. Gopar}\altaffiliation{Present address: 23, rue
  du Loess, F-67037 Strasbourg Cedex (France)}
\author{K. A. Muttalib}%
\affiliation{Department of Physics, University of Florida,
P.O. Box 118440, Gainesville, FL 32611-8440}

\author{P. W\"olfle}%
\affiliation{Institut f\"ur Theorie der Kondensierten Materie,
Universit\"at Karlsruhe, Karlsruhe, Germany}\affiliation{
Institut f\"ur Nanotechnologie, Forschungszentrum Karlsruhe,
Germany.}

\begin{abstract}

We calculate the distribution of the conductance 
$P(g)$ 
for a quasi
one-dimensional system in the metal to insulator crossover regime, based
on a recent analytical method valid for all strengths of disorder.  
We show the evolution of $P(g)$ as a function of the disorder parameter
from a insulator to a metal. Our results agree with numerical studies
reported on this problem, and with exact analytical results for the
average and variance of g.

\end{abstract}


\maketitle

\section{Introduction}

The study of electronic transport in quantum mesoscopic disordered
systems has been a topic of interest for a long time
\cite{{reviews},{datta}}. Due to the random positions of the
impurities in such systems, quantum interference effects give rise
to strong fluctuations in the conductance $g$ from sample to
sample. These effects can be observed in a single disordered
sample by changing, for example, the applied magnetic field
\cite{wash-webb,mohanty}, since this is similar to a change in the
impurity configuration of the sample. As a consequence of these
fluctuations a statistical study of the conductance is required.

Many efforts have been made in order to have a complete
statistical description of the conductance in the three different
regimes of transport \cite{reviews}: metallic ($\xi \gg L$), where
$\xi$ is the localization length and $L$ the typical size of the
system, insulating ($\xi \ll L$) and crossover ($\xi \sim L$). It
is known that in the metallic regime the distribution of the
conductance $P(g)$ is Gaussian, so the first and second moments,
i.e., the average $\langle g \rangle$ and the variance ${\rm
var}(g)$ are enough to describe $P(g)$. It turns out that in the
deeply metallic regime ${\rm var}(g)$ is a pure number independent
of the details of the system known as the universal conductance
fluctuations, depending only on the presence or absence of time
reversal symmetry and spin-rotational symmetry
\cite{altshuler-lee-stone}. In the opposite regime of transport
(insulating regime), the distribution of $g$ is log-normal, which
means that $\ln g$ follows a Gaussian distribution.

The intermediate regime of transport between the metallic and
insulating regimes can be reached, for example, by increasing the
disorder in a metallic sample. As the disorder is increased the
conductance fluctuations grow in such way that ${\rm var}(g)$
becomes of the same order as $\langle g \rangle$. In this case,
the first two moments are no longer enough to describe $P(g)$.
In fact, the full distribution of $g$ is
needed in order to have a good statistical characterization of the
electronic transport. However, not much is known about the
distribution of the conductance in the crossover regime, even in
the case of a simple geometry like a quasi one-dimensional system
or quantum wire ($L >> W$, where $L$ is the length and $W$ is the
width of the system), where a smooth transition from the metallic
to the insulating regime exists.

There are a number of numerical simulations in the intermediate
regime which show a broad asymmetric distribution of $g$ with a
flat part for values of $g < 1$ and a strong decay for $g >1$
\cite{{plerou-wang},{markos},{markos1},{ru-sou}, {ru-ma-sou}} .
Also, these numerical results have shown an interesting
qualitatively similar behavior of $P(g)$ for quasi one, two and
three dimensional systems \cite{{markos},{ru-sou},
{ru-ma-sou},{markos1}}.

With respect to analytical results in the crossover region, the
first two moments of the distribution of the conductance have been
calculated, in fact, for all values of disorder for quasi one
dimensional systems, using the supersymmetric non-linear sigma
model ($\sigma$ model) \cite{zirnbauer,mirlin-muller}. For the
complete distribution $P(g)$ beyond the metal and insulator
limits, a systematic method has recently been developed by two of
us \cite{mutt-wolfle}. Using this method, we were able to  
calculate $P(g)$ for quasi one-dimensional systems  {\it near the
crossover regime, approaching on the insulating side}. It was
predicted that on the insulating side $P(g)$ follows a ``one-side''
log-normal distribution cut off by a Gaussian for $g>1$. Numerical
calculations have shown a cutoff at $g=1$ in the crossover and
insulating regimes \cite{{ru-sou},{ru-ma-sou},{markos1}} and a
``one-side'' log-normal distribution for $g < 1$, in the
insulating region \cite{{markos1},
{garcia-saenz}}. The
method reproduces other well known insulating and metallic limits.
For example, using the results of \cite{mutt-wolfle}, Fig.
\ref{fig1} shows the average and variance of $g$ as a function of
the disorder parameter $\Gamma=\xi/L$; the first two moments agree
quantitatively well with those from the $\sigma$ model, in the
metallic and insulating limits. In Fig. \ref{fig2}, $P(g)$ is
plotted for a metal and insulator case, using again the results of
\cite{mutt-wolfle}, which shows that the two limits are well
captured by the method. However, the crossover regime is
qualitatively correct only on the metallic and insulating sides,
suggesting that the approximations made in \cite{mutt-wolfle} are
not as good for the complete crossover region. Indeed, numerical
simulations have shown that the drop in $P(g)$ at $g>1$ appears to
be exponential \cite{{ru-ma-sou},{markos1}}, as opposed to the
Gaussian cutoff predicted in \cite{mutt-wolfle}. It is therefore
important to improve the approximations made in \cite{mutt-wolfle}
in order to obtain a better description of the crossover regime.

In this work we calculate the distribution of
the conductance {\it in the crossover regime} using the
proposed method in \cite{mutt-wolfle} with some improvements.
We show how $P(g)$ 
changes from a broad highly asymmetric ``one-sided" log-normal
distribution on the insulating side of the crossover regime to a
Gaussian-like distribution on the metallic side of the crossover
regime, as function of the disorder $\Gamma$. The
distribution at the crossover regime agrees with numerical
results, except exactly at $g=1$, which may be due to the
existence of a singularity \cite{mu-wo-ga-go,markos1} which might
make our approximations less accurate.

\section{Scattering approach and DMPK equation}

In order to study the conductance, we will use the scattering
approach to electronic transport, in which the (dimensionless)
conductance  is given  in terms of the transmission eigenvalues
$T_n$ by
\begin{equation}
\label{sumT}
g=\sum_{n=1}^{N}T_n,
\end{equation}
where $N$ is the number of channels or transverse modes of the
wire. Due to the random positions of the impurities in a
disordered sample, the eigenvalues $T_n$ fluctuate from sample to
sample and the distribution of $g$ is given in general by $P(g)=
\left< \delta \left(g - \sum_{i=1}^{N}T_n \right) \right>$, where
$\langle ... \rangle$ indicates the average with the joint
probability distribution of the transmission eigenvalues
$p(\{T_n\})$.

For a mesoscopic quasi one-dimensional quantum wire, the evolution
equation for the distribution $p(\{T_n\})$ as a function of the
length of the system is given by the Dorokhov-Mello-Pereyra-Kumar
(DMPK) equation \cite{dorokhov-dmpk}. For systems without time
reversal invariance or unitary symmetry, which is the case that we
study here, the DMPK equation can be written as
\begin{equation}
\label{dmpk}
l\frac{\partial p({\bf\lambda})}{\partial
L}=\frac{1}{N}\sum_{n=1}^{N}
\frac{\partial}{\partial\lambda_n}\lambda_n(1+\lambda_n)J({\bf\lambda})
\frac{\partial}{\partial
\lambda_n}\frac{p({\bf\lambda})}{J({\bf\lambda})},
\end{equation}
where
\begin{equation}
J=\prod_{i=1}^N\prod_{j=i+1}^N \mid \lambda_j - \lambda_i
\mid ^2 ,
\end{equation}
$\lambda_n=(1-T_n)/T_n$ and $l$ is the mean free path. The solution of
the DMPK equation has been found by Beenakker and Rejaei
\cite{beenakker-rejaei} for all degree of disorder and it is given
by
\begin{eqnarray}
\label{pofxB-R}
p(\{x_i\})=C\prod_{i<j}&(&\sinh^2x_j-\sinh^2x_i)\prod_{i}\sinh
2x_i
\times \nonumber \\
&& {\rm Det}\left[\int_0^{\infty}e^{-k^2s/4N}\tanh(\pi k/2)k^{2m-1} \times
P_{(ik-1)/2}(\cosh 2x_n)  \right] ,
\end{eqnarray}
where the $x_i$'s are related with the $\lambda_i$'s by
$\lambda_i=\sinh^2 x_i$ or with the transmission eigenvalues by
$T_i=1/\cosh ^2 x_i$; $C$ is a normalization factor and
$P_{(ik-1)/2}$ is a Legendre function.

Using the Fourier representation of the $\delta$-function, the
distribution of the conductance is given in general by
\begin{equation}
\label{pofgfull}
P(g)=
\frac{1}{Z}\int_{-\infty}^{\infty}\frac{d\tau}{2\pi}\int_{0}^{\infty}
\prod_{i=1}^{N}d x_i\exp\left[i\tau\left(g-\sum_i^N
    \frac{1}{\cosh^2 x_i}\right)\right]
p\left( \{x_i\}\right)
\end{equation}
with $p\left( \{x_i\}\right)$ given by Eq. (\ref{pofxB-R}). The
calculation of $P(g)$ from Eq. (\ref{pofgfull}) involves a nontrivial
N-fold integration.

\section{Features of the current method}

In the metallic regime, where the transmission eigenvalues $T \sim
1$ contribute significantly to the conductance, the $\lambda$'s
($=(1-T)/T$) are very close to each other and a continuum density
for $\lambda$ can be assumed. In this approximation the universal
conductance fluctuations can be derived. On the other hand, in the
insulating regime, the  $\lambda$'s are much larger than the unity
(small transmission) and separated exponentially. In this case,
the lowest eigenvalue gives the most important contribution to the
conductance and within this approximation the log-normal
distribution for $g$ can be derived. But none of these
approximations can describe the crossover regime where the
eigenvalues are neither very close to each other, nor
exponentially separated. We will see, however, that it is possible
to study $P(g)$ in the crossover region by making systematic
corrections from the metal and insulating regimes.

In order to calculate the distribution $P(g)$ in the crossover
regime, we propose, following \cite{mutt-wolfle}, to {\it separate
out the two lowest eigenvalues} $x_1,x_2$, {\it treating the rest
as continuum}. In \cite{mutt-wolfle} only the lowest eigenvalue
$x_1$ was separated out. We will see that by separating out one
additional eigenvalue we can go beyond the region studied in
\cite{mutt-wolfle}. In the following we give a brief description
of the important features of the current method in order to calculate
$P(g)$.

The general solution of the DMPK equation (\ref{pofxB-R}) can be
simplified for the cases $l \ll L \ll N l$ (metal) and $l \ll Nl
\ll L$ (insulator). In these cases the integral over $k$ can be
calculated analytically and the determinant is given, for
example in the metallic regime, by a product of
the difference of
the square eigenvalues \cite{beenakker-rejaei}. The solution for
both metallic and insulating regimes can be written as:
\begin{equation}
\label{pofxmetal-insulator}
p(\{x_i\})= \frac{1}{Z}\exp\left[-H(\{x_i\}) \right]
\end{equation}
with
\begin{equation}
\label{H}
H\left( \{x_i\}\right)=\sum_{i<j}^N u(x_i,x_j)+\sum_i^N V(x_i) .
\end{equation}
The $H\left( \{x_i\}\right)$ function can be interpreted as the
Hamiltonian of N charges at the positions $x_i$ with a two body interaction
given by $u(x_i,x_j)$ and confinement potential $V(x_i)$. Those terms
are given by
\begin{equation}
\label{umetal}
u(x_i,x_j)=-\frac{1}{2}\ln
|\sinh^2x_j-\sinh^2x_i|-\frac{1}{2}\ln|x^2_j-x^2_i | ,
\end{equation}
\begin{equation}
\label{V(x)}
V(x_i)=\left\{
\begin{array}{lr}
\frac{1}{2}\Gamma x_i^2-\frac{1}{4}\ln(x_i\sinh2 x_i) & \textrm{
(metal)} \\
\frac{1}{2}\Gamma x_i^2-\frac{1}{4}\ln(x_i\sinh2 x_i) -\frac{1}{4}\ln
x_i & \textrm{(insulator)} .
\end{array}  \right.
\end{equation}
The parameter $\Gamma(=\xi/L)$  is given by $\Gamma = Nl/L$ in
quasi-one-dimension. $\Gamma \gg 1$ and $\Gamma \ll 1$ correspond
to the metallic and insulating limits, respectively.
$\Gamma \sim 1$ is the crossover regime. We can see from Eq.
(\ref{V(x)}) that the one body potential  $V(x_i)$ in the
insulating regime differs by a logarithmic term from the
corresponding $V(x_i)$ for a metal. In the localized regime the
transmission is very small, i.e. $x_i \gg 1$ (since $T=1/\cosh^2
x$)  and then the logarithmic term is negligible compared to the
other terms of $V(x_i)$. This means that the solution in the
metallic regime might be used for the insulating regime as well.
Therefore, we will assume that the solution in the metallic regime
is also valid for the intermediate regime. We remark that it is not
always possible to write the general 
solution $p(\{x_i\})$, Eq. (\ref{pofxB-R}), in the form Eqs. (6-7). 
However,  
we will check the
validity of our approximations by comparing the average and
variance of $g$ with the respective exact results from the
$\sigma$ model for those quantities \cite{mirlin-muller}.

Using Eqs. (\ref{pofgfull}) and (\ref{pofxmetal-insulator}), the
distribution of the conductance can be written as
\begin{equation}
\label{pofgN}P(g)=
\frac{1}{Z}\int_{\infty}^{\infty}\frac{\tau}{2\pi}\int_{0}^{\infty}
\prod_{i}^{N}dx_i\exp\left[i\tau\left(
g-\sum_{i}^{N}\frac{1}{\cosh ^2 x_i}\right ) - H(x_i) \right], \nonumber
\end{equation}
where $Z$ is a normalizing factor.
In order to calculate the distribution of $g$, as mentioned
before, we start by separating out the two lowest eigenvalues
$x_1$ and $x_2$ in $H(x)$, Eq. (\ref{H}):
\begin{equation}
H= H_{1,2}+\sum_{3 \leq i<j}u(x_i,x_j)+\sum_{3 \leq i}V(x_i)
\end{equation}
with
\begin{equation}
H_{1,2}=u(x_1,x_2)+ \sum_{3 \leq i}u(x_1,x_i)+
\sum_{3 \leq i}u(x_2,x_i)+V(x_1)+V(x_2)
\end{equation}
Now we make the continuum approximation:
\begin{eqnarray}
\label{Hcontinuum}
H(x_1,x_2,x_3,\sigma(x))=V(x_1)&+&V(x_2)+ u(x_1,x_2)+
\int_{x_3}^{\infty}dx\sigma(x)u(x_1,x)+
\int_{x_3}^{\infty}dx\sigma(x)u(x_2,x)\nonumber \\
&+&\frac{1}{2}\int_{x_3}^\infty dx\int_{x_3}^\infty
dx'\sigma(x)u(x,x')\sigma(x')+ \int_{x_3}^\infty dx \sigma(x)V(x) ,
\end{eqnarray}
where we have introduced the density of eigenvalues $\sigma(x)$
which has to be calculated in a self-consistent way and subject to
the conditions $\sigma(x)>0$ and $\int_{x_3}^\infty \sigma(x)dx =
N-2$.

Defining the ``free energy'' F as
\begin{equation}
\label{freeenergy}
F(x_1,x_2, x_3;\sigma(x))= 2
H(x_1,x_2,x_3;\sigma(x))+i\tau\left[\frac{1}{\cosh^2(x_1)}+
\frac{1}{\cosh^2(x_2)}
+\int_{x_3}^{\infty}\frac{\sigma(x)}{\cosh^2x}\right] ,
\end{equation}
the distribution $P(g)$ can now be written as a functional integration:
\begin{equation}
\label{pofgfunctional}
P(g)=\frac{1}{Z}\int_{\infty}^{\infty}\frac{d\tau}{2\pi}e^{i\tau g}
\int_{0}^{\infty}dx_1\int_{x_1}^{\infty}dx_2\int_{x_2}^{\infty}dx_3\int
\mathcal{D}[\sigma(x)]
e^{-F(x_1,x_2, x_3;\sigma(x))} .
\end{equation}

As in \cite{mutt-wolfle}, in order to find the density $\sigma(x)$
self-consistently, we minimize the free energy by taking the
functional derivative: $\delta  F(x_1,x_2, x_3;\sigma(x))/\delta
\sigma(x)=0$, which gives the following integral equation for the
saddle point density $\sigma_{sp}(x)$:
\begin{equation}
\label{intequation}
-\int_{x_3}^{\infty}dx'u(x,x')\sigma_{sp}(x,x')=
u(x,x_1)+u(x,x_2)+\frac{i\tau}{\cosh^2 x}+V(x) .
\end{equation}
The above integral equation can be solved when the lower limit
goes to zero. In this case, the integral can be symmetrically
extended to negative values which allows one to invert the kernel
and obtain $\sigma_{sp}(x)$. For the non-zero limit, we calculate
$\sigma_{sp}(x)$ using the following ``shift approximation''. In
terms of the eigenvalues $\lambda$ and $\lambda'$, the interaction
term in $H$, Eq. (\ref{umetal}), can be written as:
\begin{equation}
u(\lambda,\lambda')=-\frac{1}{2}\left[ u_1(\lambda,\lambda') +
u_2(\lambda,\lambda' ) \right]
\end{equation}
with
\begin{eqnarray}
u_1(\lambda,\lambda')&=&\ln |\lambda -\lambda'| , \\
u_2(\lambda,\lambda')&=& \ln \left[\left(\sinh^{-1}\sqrt{\lambda}\right)^2-
\left( \sinh^{-1}\sqrt{\lambda'} \right)^2  \right] .
\end{eqnarray}

We note that $u_1$ is translationally invariant in the variables
$\lambda,\lambda'$, but $u_2$ is only invariant in the metallic
regime ($\lambda \ll 1$). However, in the insulating regime
($\lambda \gg 1$) $u_2$ is negligible compared to $u_1$. This fact
suggests that we write $u_2$ for the shifted variables
$\eta=\lambda-\lambda_3, \eta'=\lambda'-\lambda_3 $ as
\begin{equation}
\label{u2shift}
u_2(\eta+\lambda_3,\eta'+\lambda_3)=u^0(\eta,\eta')+\Delta
u_2(\eta,\eta';\lambda_3) ,
\end{equation}
where
\begin{equation}
\Delta u_2(\eta,\eta';\lambda_3)=\ln \frac{
  (\sinh^{-1}\sqrt{\eta+\lambda_3})^2 -
  (\sinh^{-1}\sqrt{\eta'+\lambda_3})^2 }
 {(\sinh^{-1}\sqrt{\eta})^2-(\sinh^{-1}\sqrt{\eta'})^2 } .
\end{equation}
Then, the integral equation (\ref{intequation}) is written as
\begin{eqnarray}
\label{intequation2}
\int_{-\infty}^{\infty}ds \left[\ln |\sinh (t-s)| + \ln |t-s|+
\frac{1}{2}\Delta u_2(t,s) \right] \sigma_{sp}(s)&=&  u(\lambda+
\lambda_3,\lambda_1)+
u(\lambda+\lambda_3,\lambda_2)\nonumber \\&+&\frac{i\tau}{1+\lambda+\lambda_3}+
V(\lambda+\lambda_3),
\end{eqnarray}
where $\sigma_{sp}(s)d(s)=\rho_{sp}(\eta+\lambda_3)d(\eta)$ with 
$\rho_{sp}(\lambda)d(\lambda)=\sigma_{sp}(x)dx$, while
$\sinh^2 t=\eta$ and $\sinh^2 s=\eta'$; we have also
extended the last integral to negative values since
$\sigma_{sp}(s)= \sigma_{sp}(-s)$. Finally, it turns out that
$\Delta u_2(t,s)$ can be approximated by the product of the sum of
two Lorentzians whose parameters are determined by the limits $s
\to 0$, $s \to \infty$ and $s=\lambda_3$. However, a posteriori
numerical calculation showed that the contribution to the saddle
point free energy $F_{sp}(x_1,x_2, x_3;\sigma(x))$ coming from the
$\Delta u_2(t,s)$ term was negligible. So, in order to simplify
our calculations we neglect this term. Our last simplification is
related to the interaction terms in $\sigma_{sp}(x)$: we consider
only interactions between neighboring eigenvalues i.e., $x_1, x_2$
and $x_2,x_3$ and neglect interactions between $x_1$ and $x_3$. As
mentioned before, we will check our approximations by comparing
$<g>$ and var($g$) with those from the $\sigma$ model.

Since the right hand side of Eq. (\ref{intequation2}) is linear in
$\tau$ and therefore $\sigma_{sp}(s)$, the saddle point free
energy (Eq. (\ref{freeenergy})) is quadratic in $\tau$, so
$F_{sp}$ can be written as
\begin{equation}
F_{sp}=F^0+(i\tau)F'+\frac{(i\tau)^2}{2}F''
\end{equation}
and the integral over $\tau$ in Eq. (\ref{pofgfunctional}) can
be done exactly with the following result:
\begin{equation}
\label{pofglast}
P(g)=\int_{0}^{\infty}dx_1\int_{x_1}^{\infty}dx_2 \int_{x_2}^{\infty}dx_3
e^{-S},
\end{equation}
where
\begin{equation}
S= -\frac{1}{2F''}\left( g-F' \right)^2 + F^0 .
\end{equation}

Therefore once $F_{sp}$ is obtained from $\sigma_{sp}(x)$, the
calculation of the distribution $P(g)$ is reduced to a triple
integration, Eq. (\ref{pofglast}), which we compute numerically.

Note that the above described method can in principle be
implemented fully numerically, without making approximations on
$\Delta u_2(t,s)$ or $\sigma_{sp}(x)$. The analytic approach
allows us to identify the dominant terms and understand their
change with disorder. The price we pay is that the expressions for
the density and free energy are not valid for all possible values
of $x_i$ ($0 < x_i <\infty$). In fact, our expressions for
$\sigma_{sp}(x)$ and $F_{sp}$ are restricted to values of $x_3 <
5$. We will see, however, that it is enough to reach the region of
main interest: the crossover regime.

\section{Results and Conclusions}

\begin{figure}
\begin{center}
\includegraphics[angle=270,width=0.36\textheight]{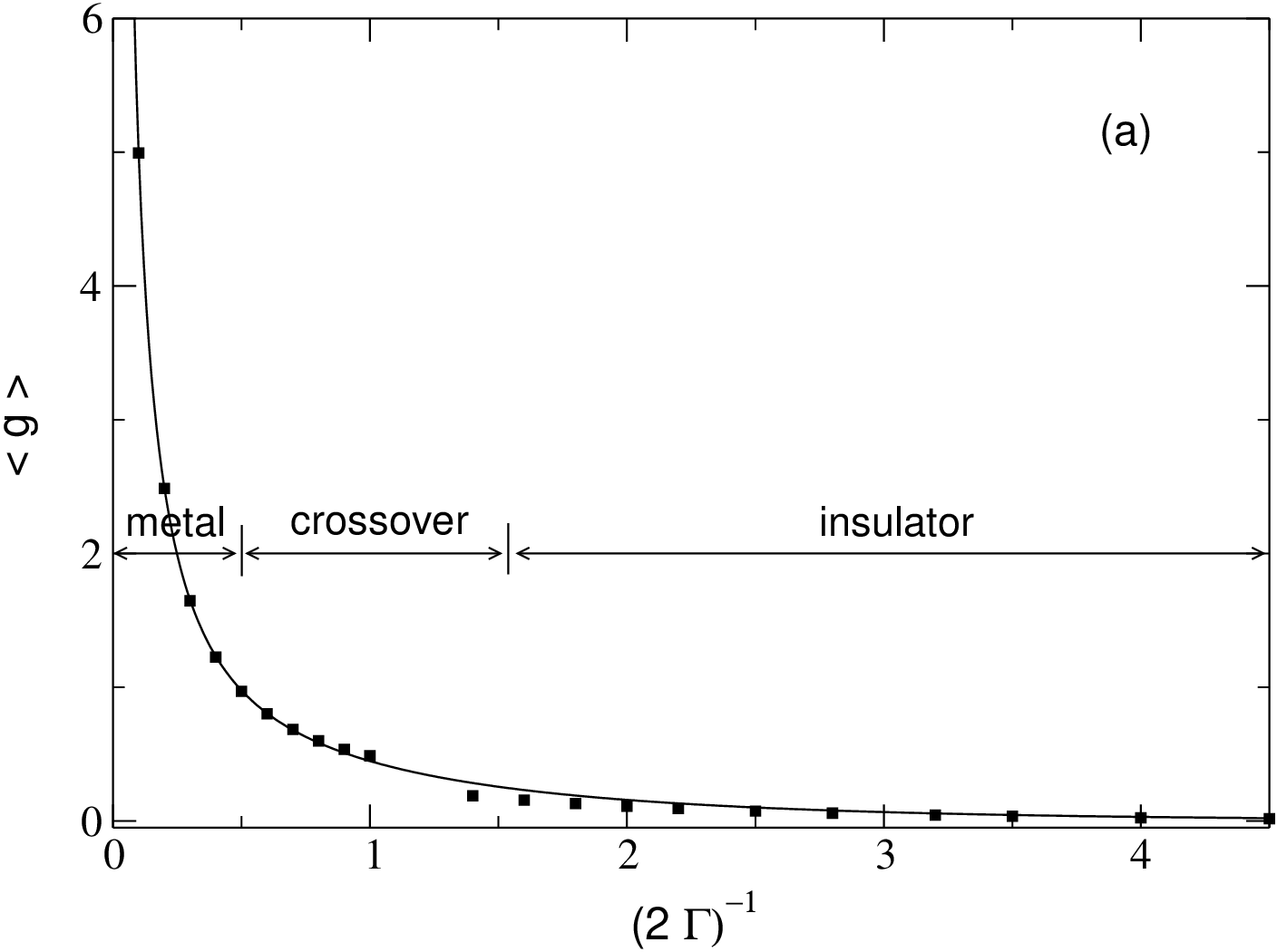}
\includegraphics[angle=270,width=0.36\textheight]{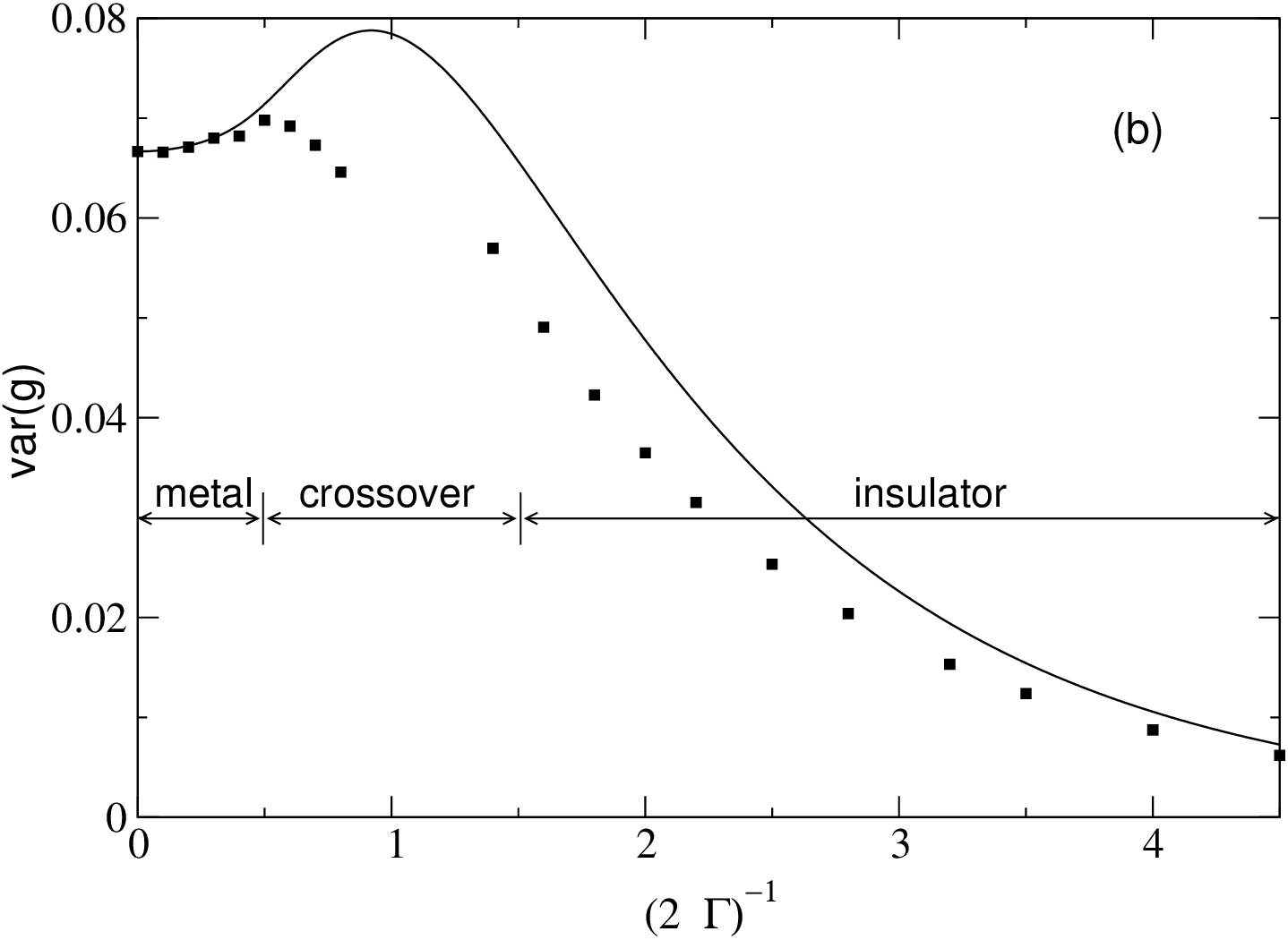}
\caption{\label{fig1} (a) Mean and (b) variance of $g$  calculated
from the results in \cite{mutt-wolfle} (squares) are compared with
the correspondent results from the
$\sigma$ model (solid line) \cite{mirlin-muller}. Horizontal
arrows show schematically the different regimes of transport. Note that
$<g>$
and var($g$) from \cite{mutt-wolfle} are not calculated for the complete
crossover region, since the deviation from the $\sigma$ model results 
becomes
large.}
\end{center}
\end{figure}

We start comparing our calculations to the first two moments
available for all degree of disorder from the $\sigma$ model
\cite{mirlin-muller}. Figure \ref{fig3} shows that the average and
variance of $g$ is now better in the crossover region, compared to
the previous calculation of \cite{mutt-wolfle}, see Fig.
\ref{fig1}. On the other hand, the metallic regime is not described
as good as in \cite{mutt-wolfle} since we have neglected 
$\Delta u_2$ in Eq. (\ref{u2shift}) and 
the interaction
terms between $\lambda_1$ and $\lambda_3$. Actually,
the approximations made in \cite{mutt-wolfle} are quite good in this metallic
regime, Fig. \ref{fig1}, and we have included the results (squares) for the average
and variance in Fig. 
\ref{fig3} for completeness. Similarly, our restriction
to values of $x_3 < 5$ for $\sigma_{sp}(x)$ does not allow us to
go throughout the insulating regime, since
in this regime the conductance is dominated by the first two
eigenvalues $x_2 \gg x_1 \gg 1$. However, in our approach this
region is very well described by two eigenvalues only as the
density goes to zero. For completeness, we also include in Fig.
\ref{fig3} the results for such a calculation. Our main goal in
this work is to develop approximations that are valid in the
crossover region, and as shown in Fig. \ref{fig3}, our current
approximations lead to very good results in the desired region.
The complete Fig. \ref{fig3} shows that the two limiting regimes
can be described within the same formulation using different set
of approximations.

\begin{figure}
\begin{center}
\includegraphics[angle=-90,width=0.36\textheight]{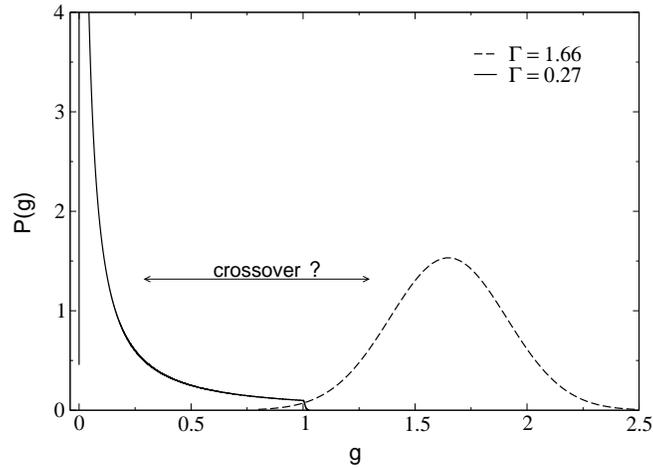}
\caption{\label{fig2} Distribution $P(g)$ in the metal (dashed line) and
insulating (solid line) regime calculated by using the results in 
\cite{mutt-wolfle}.
As expected, a Gaussian distribution is obtained for the metallic case 
while,
in a logarithmic scale, $P(\ln g)$ follows a log-normal distribution for
the insulating case.}
\end{center}
\end{figure}

\begin{figure}
\begin{center}
\includegraphics[angle=-90,width=0.36\textheight]{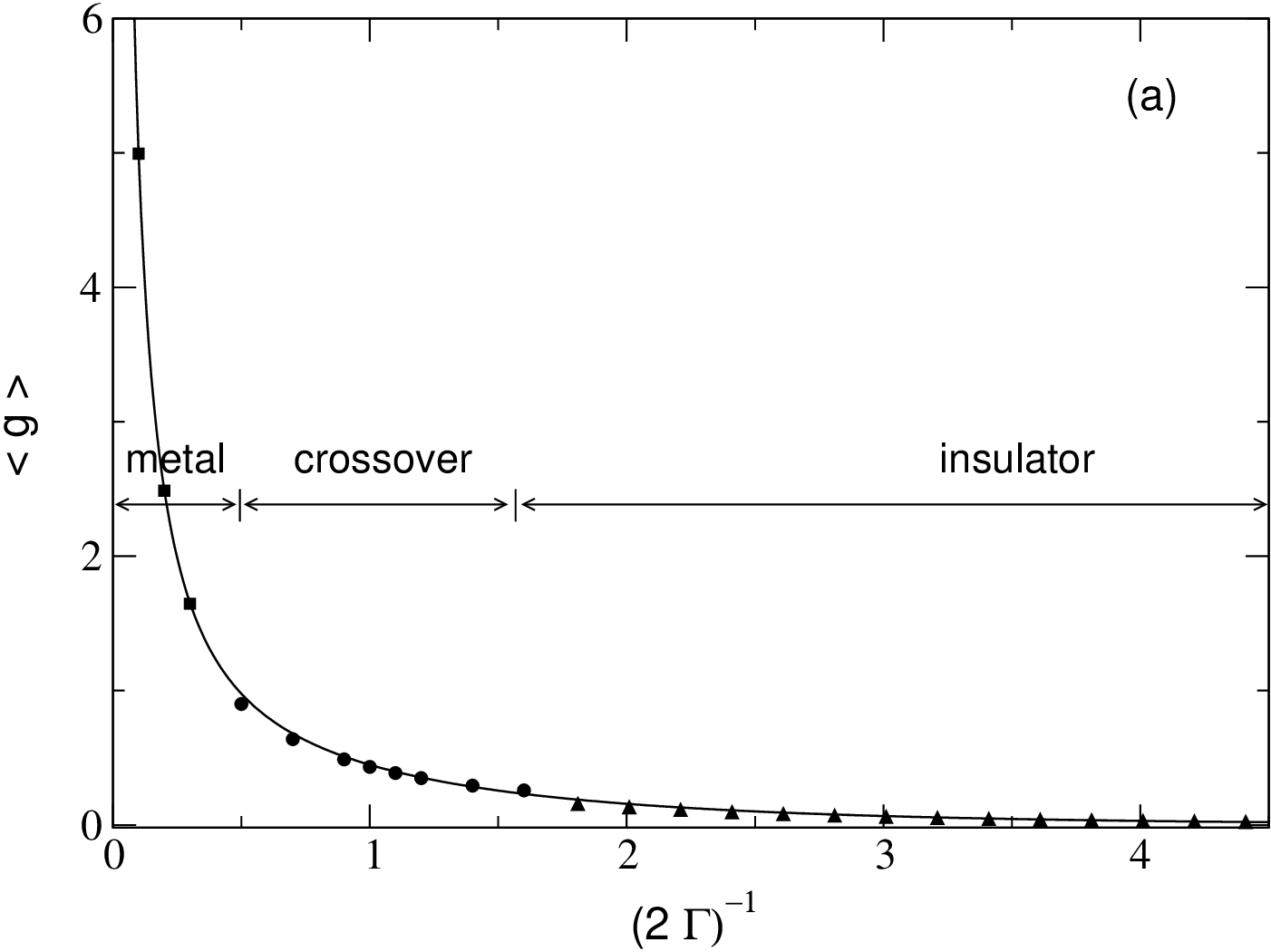}
\includegraphics[angle=-90,width=0.36\textheight]{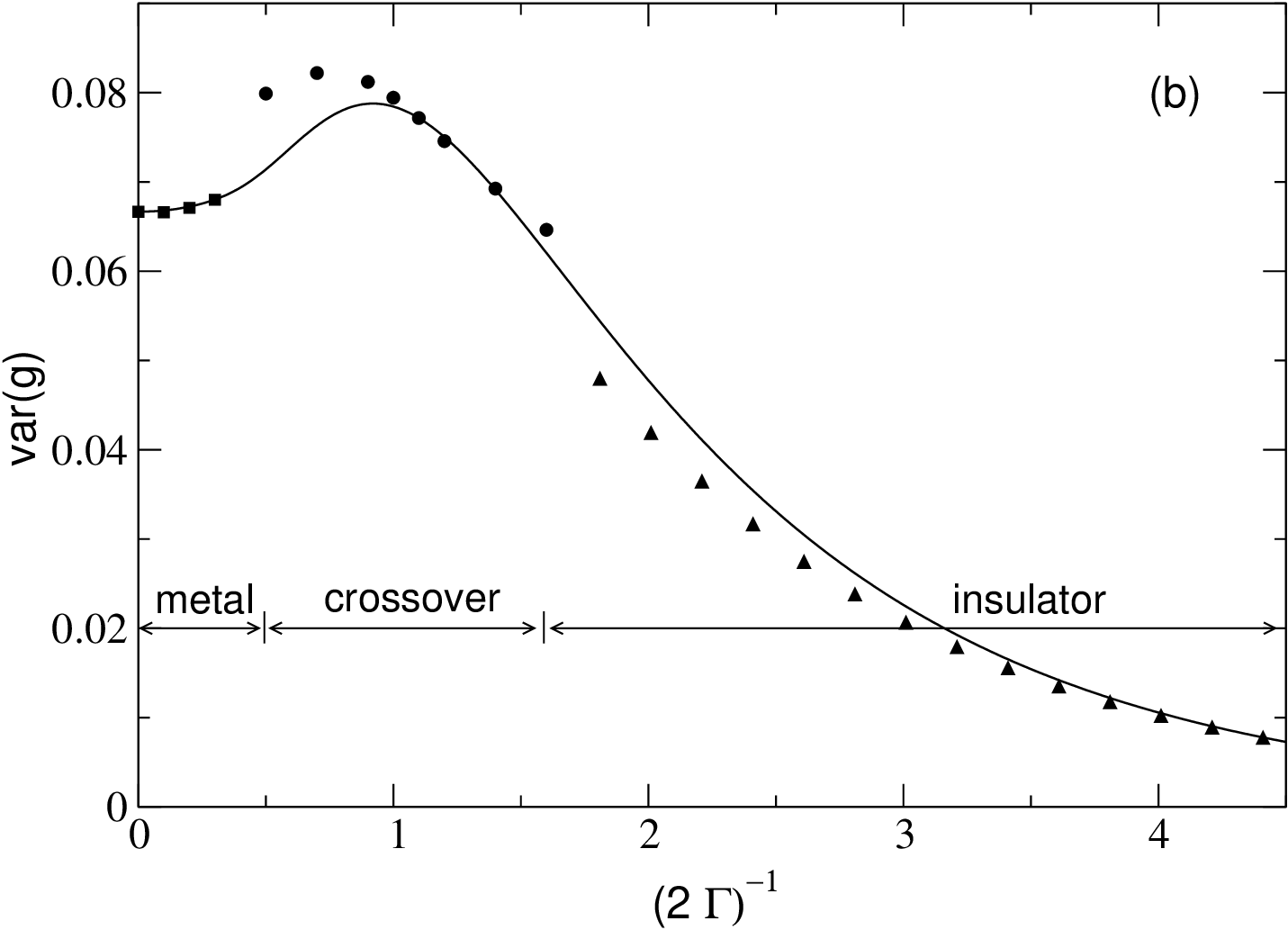}
\caption{\label{fig3} Comparison of our results for the average
and variance of the conductance, (a) and (b) respectively, with
the non linear $\sigma$ model (solid line) \cite{mirlin-muller}.
Squares are the same as in Fig. \ref{fig1}. Dots are results
obtained using the current approximations for the crossover
region. Triangles are results obtained using two eigenvalues
only.}
\end{center}
\end{figure}

\begin{figure}
\begin{center}
\includegraphics[angle=270,width=0.36\textheight]{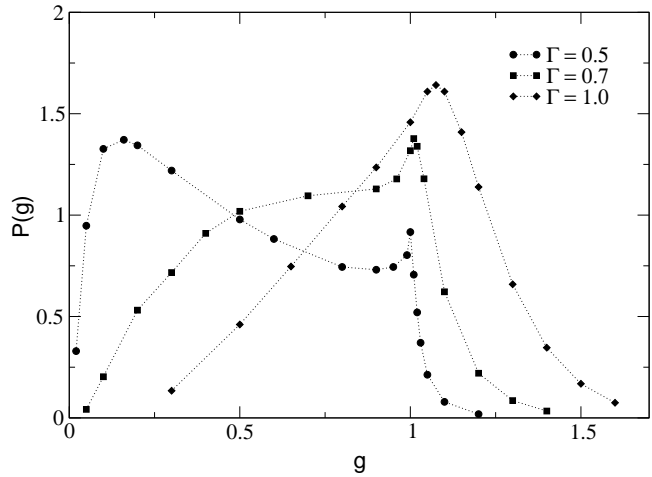}
\caption{\label{fig4} Evolution of the distribution of the conductance
  $P(g)$ as the disorder parameter $\Gamma$ is changed. Three cases are 
shown
$\Gamma=0.5$ (dots), $\Gamma=0.7$ (squares) and $\Gamma=1.0$ (diamonds).}
\end{center}
\end{figure}

In figure \ref{fig4} we show the evolution of $P(g)$ as we change
the disorder parameter $\Gamma$, evaluated within the current
approximations valid in the crossover region. Suppose that we
start decreasing the disorder in a sample i.e., we go from
insulating to metallic behavior: at $\Gamma=0.5$ we obtain a broad
distribution with a bump at small values of $g$ which is a
reminiscence of the huge peak for $g<<1$ in the insulating regime,
see Fig. \ref{fig2}. Also, at $g > 1$, $P(g)$ has an exponential
decay as has been seen in numerical simulations
\cite{{plerou-wang},{markos},{ru-sou},{ru-ma-sou}} rather than the
Gaussian cutoff at $g>1$ predicted in \cite{mutt-wolfle}. As
$\Gamma$ is increased ($\Gamma=0.7$), the bump at $g << 1$
disappears which makes $<g>$ increase and at the same time, the
decay at $g>1$ becomes smoother . Finally, when we decrease the
disorder even more ($\Gamma = 1.0$), we obtain a distribution
which starts to look like a  Gaussian distribution, as is expected
in the metallic regime.

\begin{figure}
\begin{center}
\includegraphics[angle=270,width=0.36\textheight]{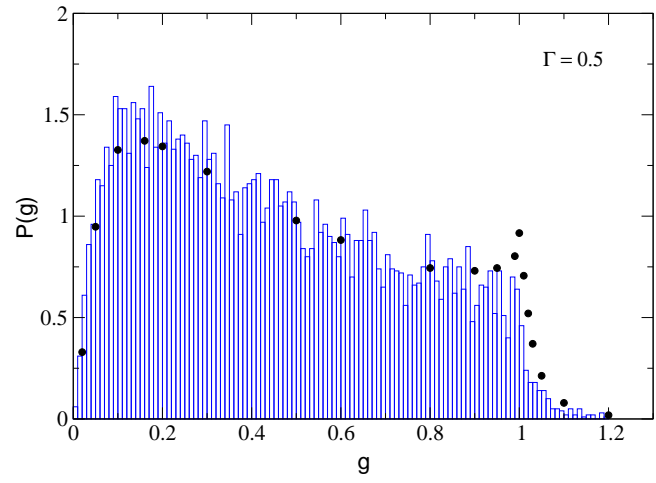}
\caption{\label{fig5} Comparison of our result for the distribution
of $g$ at $\Gamma=0.5$ (dots) with the numerical calculation presented in
(bars) \cite{plerou-wang}.}
\end{center}
\end{figure}

In order to check our results in the crossover regime,
we compare $P(g)$ at $\Gamma=0.5$ with the
distribution obtained numerically
by Plerou and Wang \cite{plerou-wang}. Figure
\ref{fig5} shows a good agreement with the numerical
simulation \cite{plerou-wang}, except at g=1 where our $P(g)$ has a
peak. Others numerical simulations \cite{{markos},{ru-sou},{ru-ma-sou},
{markos1}} do not
show any cusp at g=1  either. We believe that this behavior
at g=1 comes from our restriction $x_3 < 5$ in the density
$\sigma_{sp}(x)$, since as we go from the metallic to insulating
regime large
eigenvalues become more and more important. However, this peak eventually
develops into the
expected peak of the Gaussian in the metallic regime. On the other
hand, an essential singularity of $P(g)$ at $g=1$ has been found
in \cite{mu-wo-ga-go} in the crossover regime, on the insulating
side. So it is conceivable that the peak present for $\Gamma=0.5$
due to the existence of this singularity disappears, since the
method developed here may not be valid in the presence of such
singularities. Also, a non analytic behavior at $g=1$ has been
found numerically in quasi one and three dimensional systems in
the crossover regime \cite{markos1}.

In conclusion, we have developed a systematic method which allow
us to calculate the distribution of the conductance in a quasi
one-dimensional system. In particular we were interested in the
distribution of $g$ across the crossover regime which had not been
obtained before \cite{mutt-wolfle}. Separating out the two lowest
eigenvalues and treating the rest as a continuum in the solution
of the DMPK equation, we were able to obtain the evolution of
$P(g)$ as function the disorder parameter $\Gamma$. We have shown
how $P(g)$ develops from a broad flat distribution at the
crossover regime to a Gaussian distribution in the metallic
regime. Our results for the average and variance agree with the
exact results from the non linear sigma model and the distribution
$P(g)$ in the crossover regime agrees well with the numerical
calculation \cite{plerou-wang}, including an exponential decay for
$g>1$ reported in numerical simulations. Together with the earlier
results of how a log-normal distribution on the insulating side
develops a cutoff at $g=1$ and eventually becomes an asymmetric
`one-sided' log-normal distribution near the crossover regime,
this provides a relatively complete picture of how a log-normal
distribution becomes a Gaussian via a highly asymmetric
intermediate one-sided log-normal distributions when disorder is
changed. The quantitative details at $g=1$ at the crossover point
remains an open question.

Although our results are valid strictly only for quasi
one-dimensional systems where the DMPK equation is valid (however,
it has been shown that the restriction $L \gg W$ can be relaxed
\cite{beenakker-melsen}), a qualitatively similar behavior of
$P(g)$ is found numerically in 2 and 3 dimensional systems in the
crossover regime \cite{{markos},{markos1}}. It is therefore
possible that the method presented here could help understanding
the generic behavior in higher dimensions at the crossover regime.

\begin{acknowledgments}
We are grateful to Ziqiang Wang for sharing his results for the
distribution of the conductance in the crossover regime. One of 
the authors (V.A.G) acknowledges financial support from CONACYT,
M\'exico. KAM is grateful for hospitality at the University of Karlsruhe
and acknowledges support from SFB 195 of the DFG. PW acknowledges support
through a Max-Planck Research Award.

\end{acknowledgments}

\end{document}